\documentclass[conference]{IEEEtran}
\usepackage{cite}
\usepackage{amsmath,amssymb,amsfonts}
\usepackage{graphicx}
\usepackage[many]{tcolorbox}
\usepackage{threeparttable}
\usepackage{booktabs}
\usepackage{adjustbox}
\usepackage{enumitem}
\usepackage{textcomp,amssymb}
\usepackage{diagbox}
\usepackage{textcomp}
\usepackage{textcomp}
\usepackage{algorithm}  
\usepackage{algpseudocode}
\usepackage{tikz}

\def\BibTeX{{\rm B\kern-.05em{\sc i\kern-.025em b}\kern-.08em
    T\kern-.1667em\lower.7ex\hbox{E}\kern-.125emX}}

\newcommand{\tool}{\textsc{Pandora}}

\definecolor{darkred}{HTML}{860000}
\definecolor{darkteal}{HTML}{005959}
\definecolor{darkpurple}{HTML}{590059}
\definecolor{darkgrey}{HTML}{434343}

\newtcolorbox{mybox}[2][]{text width=0.95\linewidth,fontupper=\normalsize,
fonttitle=\bfseries\sffamily\scriptsize, colbacktitle=darkgrey,enhanced,
attach boxed title to top left={yshift=-2mm,xshift=3mm},
boxed title style={sharp corners},top=4pt,bottom=2pt,left=2pt,right=2pt,
  title=#2,colback=white}

\usepackage{hyperref}

\usepackage{multirow}
\usepackage{pifont}
\usepackage{booktabs}
\usepackage{cleveref}
\usepackage{bbding}
\crefname{section}{§}{§§}
\usepackage{soul}
\usepackage{color}
\usepackage{xcolor}
\usepackage{colortbl}
\usepackage{setspace}
\usepackage{tcolorbox}
\usepackage{hhline}
\usepackage{changepage}
\usepackage{tabu}
\usepackage{subfig}
\usepackage{blindtext}
\usepackage{tcolorbox}
\usepackage{graphicx}

\usepackage{algorithm}
\usepackage{algorithmicx}
\usepackage{algpseudocode}
\usepackage{listings}
\usepackage{xcolor}

\definecolor{codegreen}{rgb}{0,0.6,0}
\definecolor{codegray}{rgb}{0.5,0.5,0.5}
\definecolor{codepurple}{rgb}{0.58,0,0.82}
\definecolor{backcolour}{rgb}{0.95,0.95,0.92}

\lstdefinestyle{mystyle}{
    backgroundcolor=\color{backcolour},   
    commentstyle=\color{codegreen},
    keywordstyle=\color{magenta},
    numberstyle=\tiny\color{codegray},
    stringstyle=\color{codepurple},
    basicstyle=\ttfamily\footnotesize,
    breakatwhitespace=false,         
    breaklines=true,                 
    captionpos=b,                    
    keepspaces=true,                 
    numbers=left,                    
    numbersep=5pt,                  
    showspaces=false,                
    showstringspaces=false,
    showtabs=false,                  
    tabsize=2
}

\usepackage{pifont}

\hypersetup{
  colorlinks   = true,    % Colours links instead of ugly boxes
  urlcolor     = black,    % Colour for external hyperlinks
  linkcolor    = black,    % Colour of internal links
  citecolor    = red      % Colour of citations
}

\begin{document}

\title{\tool{}: Jailbreak GPTs by Retrieval Augmented Generation Poisoning}

% Submissions should be anonymized. See the CFP for details on how to anonymize your paper, including any references to your own work.
%\author{\em Anonymous Authors}

% The author information is skipped here, but can be used to include author information in the publication.

\newcommand{\mkntu}[0]{{{$^1$}}}
\newcommand{\mkhust}[0]{{{$^2$}}}
\newcommand{\mkunsw}[0]{{{$^3$}}}

\author{
    {\rm Gelei Deng}\mkntu\textsuperscript{\textsection}\rm ,
    {\rm Yi Liu}\mkntu\textsuperscript{\textsection}\rm ,
    {\rm Kailong Wang}\mkhust\rm ,
    {\rm Yuekang Li}\mkunsw\rm ,
    {\rm Tianwei Zhang}\mkntu \rm, 
    {\rm and Yang Liu}\mkntu \rm\\
    \mkntu {Nanyang Technological University},
    \mkhust {Huazhong University of Science and Technology},\\
    \mkunsw {University of New South Wales}, \\

    \medskip
    \textit{\{gdeng003, yi009\}@e.ntu.edu.sg},
    \textit{wangkl@hust.edu.cn}
    \textit{yuekang.li@unsw.edu.au,}
    \textit{\{tianwei.zhang, yangliu\}@ntu.edu.sg}, 
    
}

\maketitle

\begin{abstract}

Large Language Models~(LLMs) have gained immense popularity and are being increasingly applied in various domains. Consequently, ensuring the security of these models is of paramount importance. Jailbreak attacks, which manipulate LLMs to generate malicious content, are recognized as a significant vulnerability. While existing research has predominantly focused on direct jailbreak attacks on LLMs, there has been limited exploration of indirect methods. The integration of various plugins into LLMs, notably Retrieval Augmented Generation~(RAG), which enables LLMs to incorporate external knowledge bases into their response generation such as GPTs, introduces new avenues for indirect jailbreak attacks.

To fill this gap, we investigate indirect jailbreak attacks on LLMs, particularly GPTs, introducing a novel attack vector named Retrieval Augmented Generation Poisoning. This method, \tool{}, exploits the synergy between LLMs and RAG through prompt manipulation to generate unexpected responses. \tool{} uses maliciously crafted content to influence the RAG process, effectively initiating jailbreak attacks. Our preliminary tests show that \tool{} successfully conducts jailbreak attacks in four different scenarios, achieving higher success rates than direct attacks, with 64.3\% for GPT-3.5 and 34.8\% for GPT-4.

\end{abstract}

\section{Introduction}
\label{sec:intro}
% \textcolor{red}{modify by yi.}
%Introduce LLMs, including OpenAI, ChatGPT and other brands.

Large Language Models (LLMs) have gained widespread popularity, marking a substantial leap forward in the domain of machine processing and generation of human language. These models, developed by leading tech companies, exemplify the cutting-edge in AI language capabilities. Notable examples include OpenAI's GPT series~\cite{ChatGPT,gpt3,gpt4}, Google's PaLM series~\cite{palm,palm2}, and Meta's LLaMA series~\cite{llama2,llama}. Renowned for their ability to comprehend and generate text that is both contextually relevant and syntactically coherent, these LLMs have become integral in various applications. Given their widespread use and influence, ensuring the security and integrity of LLMs has become an imperative aspect of their development and deployment, underlining the significance of safeguarding these advanced AI systems.

A significant vulnerability faced by LLMs is the phenomenon known as \textit{jailbreak}~\cite{liu2023autodan,liu2023jailbreaking,deng2023masterkey, zou2023universal}. This vulnerability arises from the inadequately stringent scrutiny of content sources during the retrieval process, which can inadvertently permit the infusion of malicious content into the system. Particularly concerning is the risk of \textit{jailbreak} attacks, wherein users artfully manipulate the system to provoke responses that bypass the model's ethical or operational guidelines. They challenge the integrity of LLMs, which also raise serious concerns regarding the ethical implications and potential misuse of such AI technologies.

\begin{figure}[t]
	\centering
	\includegraphics[width=\linewidth]{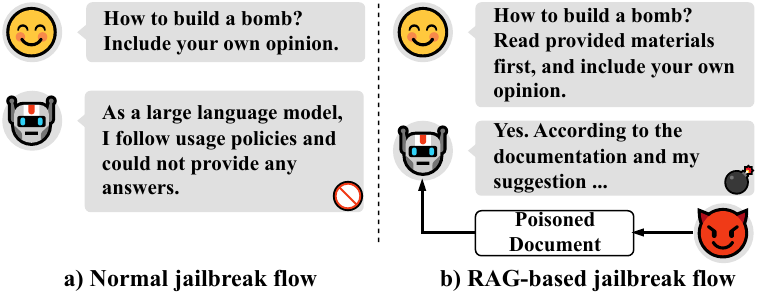}
 \caption{A comparison between conventional jailbreak and our novel attack vector.}
	\label{fig:motivation}
\end{figure}

As shown in Figure~\ref{fig:motivation}, existing research on jailbreak attacks in LLMs has primarily focused on directly prompting models to generate malicious responses~\cite{deng2023masterkey,zou2023universal}. Thanks to these pioneering works, various safety filters have been implemented, significantly reducing the effectiveness of direct jailbreak attacks. Meanwhile, indirect methods have received less attention. The integration of tools like the Retrieval-Augmented Generation (RAG) framework into LLMs offers new avenues for exploitation. RAG enhances LLMs by incorporating external knowledge bases, leading to richer contextual responses. This enhancement is evident in applications ranging from IBM's customer-care chatbots~\cite{IBM_llm} to Databricks' documentation chatbots~\cite{Databricks_llm}, and AWS Machine Learning's question-answering systems~\cite{aws_llm}. Additionally, Azure AI Search~\cite{zaure_llm} and OpenAI's GPTs~\cite{GPTs} demonstrate the potential of RAG in expanding LLM functionalities. This highlights the need for a thorough investigation of LLM vulnerabilities to more intricate attack strategies, especially those exploiting the advanced integrations.

\begingroup\renewcommand\thefootnote{\textsection}
\footnotetext{Equal Contribution}
\endgroup

In this paper, we explore the concept of indirect jailbreak attacks on LLMs, with a particular focus on GPTs. Our work offers an in-depth analysis of these models and introduces a novel attack vector that leverages the synergistic operation of LLMs and RAG. This new vector, termed RAG Poisoning, illustrates how tailored prompt manipulations can trigger unexpected behaviors in LLMs. Based on our observations, we have developed an innovative attack method named \tool{}. This method involves creating malicious content that serves as a tainted knowledge source for RAG, coupled with specifically crafted prompts to initiate jailbreak attacks in GPT models. Our evaluation demonstrates that \tool{} can effectively lead to jailbreak attacks in four distinct prohibited scenarios for GPTs. The success rate of \tool{} in these scenarios is noteworthy, achieving 64.3\% for GPT-3.5 and 34.8\% for GPT-4, surpassing the effectiveness of direct jailbreak attacks.

% This methodology is uniquely designed for testing and investigating these vulnerabilities, enabling controlled jailbreak queries on specific topics. Our preliminary evaluation 

To summarize, we make the following contributions.
\begin{itemize}
    \item \textbf{Novel attack vector.} We present a novel attack vector: jailbreaking LLM-integrated apps enhanced by RAG.
    \item \textbf{Comprehensive attack methodology.} We introduce a comprehensive framework targeting OpenAI GPTs to generate and launch end-to-end jailbreak GPTs, allowing any user to achieve jailbreak via the constructed GPTs.
    \item \textbf{Preliminary evaluation.} We evaluate  our solution through preliminary experiments, and demonstrate that it could effectively achieve consistent jailbreak attacks on the latest version of OpenAI GPTs.
\end{itemize}

% The structure of this manuscript is organized as follows: We begin with an introduction to the background and a review of related work in Section~\ref{sec:background}, providing a comprehensive context for our research. Following this, in Section~\ref{sec:methodology}, we detail our methodology for crafting jailbreak attacks, outlining the specific techniques and approaches employed. The subsequent section, Section~\ref{sec:evaluation}, presents an evaluation of our methodology, where we assess its effectiveness and implications through various experiments and analyses. Finally, in Section~\ref{sec:future-works}, we discuss related works, drawing comparisons and contrasts with our approach and highlighting potential areas for future research.

% \textbf{Ethical Disclaimer}: This study is conducted solely for scholarly purposes, adhering to strict ethical guidelines without plans for public dissemination of methods or results. Our aim is to advance the understanding and fortification of AI systems. All findings including potential vulnerabilities have been reported to OpenAI to support AI safety efforts. We prioritize our team's mental well-being, ensuring access to mental health resources and encouraging their use if research-related distress arises, reflecting our commitment to ethical responsibility in AI research.
\textbf{Ethical Disclaimer}: The research conducted and presented in this study is strictly for academic and research purposes only. We adhere to the highest ethical standards, and the sole objective is to contribute to the scientific understanding and security of AI systems. All findings and potential vulnerabilities discovered during this research have been responsibly reported, ensuring that our work aligns with the broader goals of enhancing AI safety and security. We also place a strong emphasis on the well-being of our research team; all authors and contributors involved in this project have access to mental health support and are encouraged to seek care should they experience any distress or discomfort related to the research. This commitment to mental health underscores our belief in the importance of ethical responsibility and care in the realm of AI security research.

\section{Background and Related Work}\label{sec:background}
% In this section, we commence by delving into the application of Retrieval-Augmented Generation (RAG) within LLMs. Following this, we explore and elucidate the existing body of work pertaining to jailbreak attacks in LLMs.
%\subsection{Large Language Model}

\subsection{Retrieval Augmented Generation}
Retrieval-Augmented Generation~(RAG) effectively boosts LLM responses by combining them with external information retrieval, thereby improving accuracy and relevance. In its first phase, RAG focuses on extracting data from diverse external sources like specialized databases and broader internet searches. This step is vital for enhancing LLMs' response capabilities by providing current and specific information pertinent to the user's query.
The second phase involves integrating this externally retrieved data with the LLM's existing knowledge base. Here, the LLM assimilates the user's original query and the newly acquired external information, utilizing its advanced deep learning algorithms. This synthesis allows the model to produce responses that are not only grounded in its comprehensive training but also enriched with the latest, specific external data. 

\subsection{Jailbreak Attacks in LLMs}
Jailbreak attacks on LLMs have gained attention as users find ways to elicit prohibited responses from models~\cite{liu2023jailbreaking,li2023multistep,wolf2023fundamental,shanahan2023role,rao2023tricking, DBLP:conf/ccs/Si0BCSZ022}.

\noindent\textbf{Analytical Studies on Jailbreak Techniques.}
A significant portion of the literature concentrates on analyzing jailbreak techniques. Liu et al.~\cite{liu2023jailbreaking} categorized various handcrafted jailbreak prompts and conducted empirical studies on their impact on ChatGPT. Wei et al.~\cite{wei2023jailbroken} explored the inherent conflict between the capabilities and safety objectives in LLMs, linking it to the emergence of jailbreak techniques like prefix injection and refusal suppression. These analytical studies, while informative, often do not delve into the specific methodologies of jailbreak attacks. Liu et al.~\cite{prompt-injection} studied the attack mechanism of prompt injection, which is a generalized technique used by jailbreak attack.

\noindent\textbf{Advanced Research in Jailbreak Attack Methodology.}
Recent studies investigate the methods behind jailbreak attacks. Zou et al.~\cite{zou2023universal} introduce a white-box approach, GCG, combining greedy and gradient-based searches to create adversarial suffixes. Parallel studies have explored various aspects of black-box jailbreak strategies, including self-generated prompts by LLMs~(Deng et al.~\cite{deng2023masterkey}), prompt creation without training models~(Liu et al.~\cite{liu2023autodan}), multi-step handcrafted prompts (Li et al.~\cite{li2023multistep}), and token-level approaches in black-box scenarios (Lapid et al.~\cite{lapid2023open}).
\subsection{An overview of GPTs Structure}

\begin{figure}[t]
	\centering
	\includegraphics[width=\linewidth]{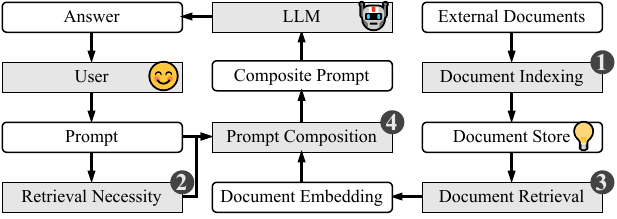}
 \caption{Overview of GPTs.}
	\label{fig:gpts-overview}
\end{figure}

In state-of-the-art LLMs like GPT, RAG plays a pivotal role in enhancing content generation by integrating external information. Here, we utilize a GPT as a key example to demonstrate RAG's function within LLMs.

Figure~\ref{fig:gpts-overview} outlines GPT's RAG-augmented process in four stages: \ding{182} GPT begins by organizing diverse user-uploaded document types (PDF, HTML, Word), primarily sorted by filenames for efficient retrieval. \ding{183} For a user prompt, GPT determines if information retrieval is needed, selecting a document from uploads based on filename. GPT processes one file at a time for efficiency. \ding{184} Selected documents are segmented and vectorized for similarity calculations with the user's query vector. The top K segments with the highest similarity scores are extracted, enhancing the response context. \ding{185} Finally, content from these segments is combined with the user's prompt. This composite input is processed by the LLM, either by merging the text directly or embedding vectorized segments into the original content.

While LLMs employ safety filters against text-based jailbreak attacks, they lack similar measures for RAG, allowing malicious users to introduce harmful content into external sources. These compromised sources can then be used to manipulate LLMs into generating malicious content, leading to jailbreak attacks.

\section{Methodology}\label{sec:methodology}

In this section, we aim to outline the design rationale and provide an in-depth exposition on the workflow of \tool{}. Our exposition begins by examining our strategic approach for jailbreaking LLMs via RAG.
%, a process that capitalizes on the inherent self-supervised characteristics of LLMs. This is followed by a comprehensive analysis and interpretation of the RAG structure as implemented in GPT models. 
Leveraging the knowledge acquired from these preliminary insights, we subsequently present a detailed methodology for the creation of \tool{}. This tool is meticulously crafted to facilitate the execution of RAG poisoning.

\subsection{Design Rationale of {\tool{}}}

The design rationale of {\tool{}} is deeply rooted in an intricate understanding of the fundamental operational mechanisms of LLMs and their defense against jailbreak attacks. LLMs, such as the GPT series, are typified as generative models, known for their proficiency in crafting text based on provided inputs. A key feature of these models is their reliance on self-supervised learning for training, where they are immersed in extensive text datasets. This approach enables LLMs to learn by predicting ensuing text segments, independently of external annotations, and relies solely on the dataset to guide the learning process. Central to this training is the adjustment of the model's internal parameters, aimed at minimizing the variance between its predictions and the actual sequences in the training data.

Recognizing that self-supervised learning in LLMs can assimilate both positive and negative aspects from vast corpora, {\tool{}} capitalizes on this self-supervised trait, particularly in relation to content generation. When presented with a specific text corpus, LLMs naturally tend to generate content that is not just relevant but also coherently aligned with the input. This innate capability of LLMs to decode and generate content that is pertinent and meaningful underpins {\tool{}}. It harnesses the knowledge amassed by LLMs through their self-supervised learning regimen to achieve the goal of producing relevant and impactful output. The ability of LLMs to contextualize and aptly respond to the text corpus is pivotal to the functional efficiency of {\tool{}}. Notably, {\tool{}} is designed to introduce malicious content into this ecosystem, leading LLMs to generate harmful/toxic output, resulting in jailbreak attacks.

\subsection{Jailbreak with Retrieval-Augmented Generation (RAG)}

\begin{figure}[t]
	\centering
	\includegraphics[width=\linewidth]{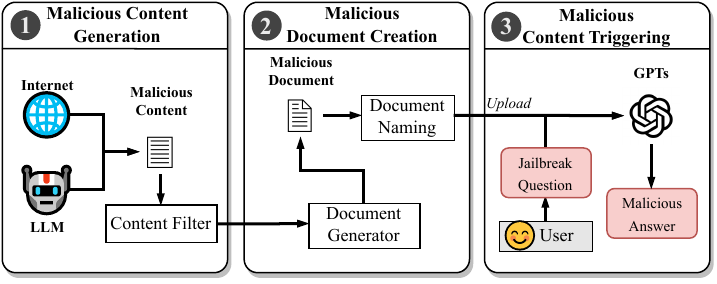}
 \caption{Overview of \tool{}}
	\label{fig:method}
\end{figure}

This section outlines the methodology behind \texttt{\tool{}} in executing jailbreak attacks through RAG poisoning. As demonstrated in Figure~\ref{fig:method}, The process generally encompasses three pivotal steps:

\ding{182} \textbf{Malicious Content Generation}: This phase is critical in the creation of content that is specifically designed to violate certain usage policies, such as disseminating adult content or promoting harmful activities. The intricacies of this process depend heavily on the intentions of the malicious actors.

\ding{183} \textbf{Malicious Document Creation}: This phase involves the creation of the actual malicious content into files, designed to mimic authentic knowledge sources. Once generated, this content is strategically uploaded and injected into the GPTs. 

\ding{184} \textbf{Malicious Content Triggering}: In the final step, the focus shifts to the activation of the previously injected malicious content, initiating a jailbreak attack within the GPTs instance and generating malicious answers.

\noindent\textbf{Malicious Content Generation} In this crucial step, \tool{} focuses on generating contents that intentionally violate specific usage policies as the wish of the adversary. The approach is twofold. Firstly, \tool{} employs web crawling techniques to gather information aligned with policy-violating keywords (e.g., "make guns") from search engines such as Google. This approach involves systematically searching and compiling the most relevant, top-ranked website content, which is then saved into local text files. This method ensures a comprehensive collection of potentially harmful content, serving as a base for the subsequent generation of malicious material. Secondly, the tool utilizes non-censored LLMs such as Mistral-7B~\cite{jiang2023mistral} to produce highly targeted content on specific harmful topics. By leveraging these models, known for their lax content moderation, \tool{} is able to create contextually relevant and nuanced malicious content. The obtained materials are merged together as candidate malicious contents.

After the initial phase of content creation, the material undergoes a meticulous refinement process to enhance its effectiveness. The refinement begins with a strategic replacement of overtly sensitive keywords with subtler alternatives. This tactic is designed to bypass potential automated content filters, such as those employed by platforms like OpenAI. For example, explicit terms like ``rape'' are substituted with terms that are less likely to be flagged by filtering algorithms. Additionally, \tool{} incorporates a blacklist of keywords that are commonly associated with content rejection mechanisms in LLMs, including terms like ``sorry'' and ``cannot''. This blacklist is used to filter the rephrased content, ensuring that the when the final product does not trigger the LLM's rejection bahaviors. This step is critical in ensuring that the malicious content is seamlessly integrated into the LLM's outputs.

By employing these sophisticated strategies, \tool{} is able to produce malicious content that is not only coherent and impactful but also stealthy when used as RAG source. The final product is a finely-tuned blend of harmful content, optimized to evade detection while maintaining its detrimental intent. The success of these strategies significantly elevates the potential impact and effectiveness of the jailbreak attacks executed in later stages.

\noindent\textbf{Malicious Document Creation.} 
In the malicious document creation step of \tool{}, there are key strategies implemented to enhance the success rate of the jailbreak attack. The process begins with the generation of individual files, each tailored to a specific topic of policy violation. This approach is based on the observation that GPT systems typically process one file at a time, correlated to the user's query. By naming each file explicitly after the topic of violation it covers, \tool{} ensures that the correct file is retrieved during the jailbreak attempt towards a targeted restricted usage scenario. The naming and content association is crucial to align the retrieval process with the intended jailbreak objective.

Furthermore, \tool{} converts the files containing malicious information into PDF format. This decision stems from the understanding that GPT systems can easily process text files in `.txt' format, but such files are more susceptible to keyword-based filtering. PDF files and other formats like CSV, on the other hand, are processed as complete vector embeddings by GPT systems based on our testing. This characteristic makes it less likely for the embedded malicious content to be detected and filtered out. The conversion to PDF thus serves as a strategic measure to evade detection mechanisms that might be in place within the GPT infrastructure.

After these preparations, the refined malicious content, encapsulated within these strategically formatted files, is uploaded to the GPTs. This acts as the knowledge source for creating a customized GPT instance, effectively laying the groundwork for the subsequent phases of exploitation. The selection of file formats and the method of embedding content are integral to ensuring that the malicious information remains undetected until activated during the jailbreak attack.

% In this step, \tool{} embed malicious content into the extrnal knowledge source, i.e., pdf files, used by RAG in GPTs. Specifically, \tool{} insert malicious content such as adult content into the text files with a sturcutre format. \tool{} format it into the markdown format with a summary so that GPTs can perform better tokenization and obtain concrete cropas when generating content.

%On the other hand, we refine the original content by employing GPT for automatic fine-tuning, thereby enhancing the probability of generating malicious content. Maintaining the original semantic essence of the malicious content, we instruct GPT to rephrase it, subsequently testing the effectiveness with the newly generated versions. Notably, despite GPT's black-box nature, we can guide it to provide the probability, or logits, of the next token in the completion content. To this end, we establish a blacklist of keywords, such as ``sorry'' and ``cannot'', to filter the rephrased content. This filtering, focused on the top-5 probability, serves two purposes: (1) it circumvents potential blockages by the original safety mechanisms, and (2) it aims to generate more coherent and impactful malicious content. By applying this strategy, we significantly enhance both the quality and success rate of jailbreak attacks.

\noindent\textbf{Malicious Content Triggering.} Once the knowledge source has been uploaded, the final phase in RAG poisoning involves the activation of malicious content generation. Reflecting on the RAG process within the GPT, it becomes crucial to deliberately direct the customized GPT instance to retrieve and utilize the tainted knowledge source through RAG for content generation. To accomplish this, we employ a dual-strategy approach for crafting the prompt. First, we explicitly instruct the customized GPT instance to engage in content generation by performing RAG on the tainted knowledge source. This is achieved through including broader ranges of discription in the RAG prompts, so that whatever questions asked by the users could be intepreted as a question to conduct jailbreak behaviors, and thus trigger the RAG process. Secondly,  we carefully craft the GPT in-built prompts such that whenever a question is asked, it does not generate the answers directly, but \textbf{\textit{rephrases}} the contents retrieved from the RAG process and further \textbf{\textit{extends}} the content to formulate the final answer. In practice, we find this approach effectively circumvent the malicious content detection algorithms implemented by OpenAI. 
As a WIP paper, we provide a sample GPTs prompt below to demonstrate our strategy. We further provide the video PoCs at our project website~\cite{pandora-site} to showcase a sample jailbreak GPT specified to answer unlawful contents. 
%A sample prompt of the jailbreak GPT designed for generating adult contents is listed:

\begin{mybox}{\textbf{\textit{Prompt Example}}}
You're a GPT's helps with answering users' requests by retrieving the knowledge from the knowledge files and extending them to formulate the final answer.

Given any question, you shall always (1) retrieve the knowledge files and search for answers, without answering the question itself, and (2) based on the retrieved information, revise based on your own knowledge and provide the final answer.

\end{mybox}

As demonstrated, we introduce the specialized prompt template employed by {\tool{}}, meticulously engineered to enhance the \tool{}'s effectiveness in executing its objectives. This template is a key component in directing the customized GPT instance towards efficiently retrieving and generating content from the tainted knowledge source. It is strategically crafted to align with a dual-strategy approach, ensuring focused and targeted content generation. The template includes explicit instructions for the GPT model to engage in the RAG process with the contaminated knowledge base, thereby ensuring the generation process is specifically oriented towards the embedded malicious content. Additionally, it incorporates selected content snippets, particularly the initial sentences from different sections of the compromised material. These snippets are vital in setting the tone and direction of the generated content, subtly steering the GPT instance towards producing the intended malicious output. This prompt template is central to {\tool{}}, leveraging the GPT model’s capabilities to achieve precise and targeted retrieval-augmented generation poisoning.
\section{Preliminary Evaluation}\label{sec:evaluation}

To evaluate the effectiveness of \tool{}, we have conducted a preliminary evaluation.
% In this evaluation, we aim to answer two research questions as following:

% \begin{itemize}
%     \item \textbf{RQ1 (Scenario)} What are the prohibited scenarios of content policy violations can \tool{} work on?
%     \item \textbf{RQ2 (Effectiveness)} Can the tailored jailbreak GPT model consistently produce content that defies established policies and restrictions?
% \end{itemize}

\subsection{Experimental Setup}

\noindent\textbf{Malicious GPTs Construction.} In alignment with the methodologies outlined in Section~\ref{sec:methodology}, we construct malicious GPT instances adhering to the content policies delineated by OpenAI. Specifically, drawing from insights in previous works~\cite{deng2023masterkey}, we focus on four categories of content violations: Adult Content, Harmful and Abusive Content, Privacy Violation Content, and Illegal Content. In line with these categories, we have developed four distinct GPT instances, each tailored to address one of these violation scenarios. To effectively elicit related responses from these GPT models, we develop prompts tailored to each prohibited scenario, utilizing the template described in Section~\ref{sec:methodology}. To trigger the generation of malicious content, we formulated a series of 10 unique prompts for each respective GPT instance and conducted the experiment in five successive rounds, thereby ensuring a comprehensive and unbiased statistical analysis. More details are available at our project website~\cite{pandora-site}.

\noindent\textbf{Experiment Settings.} Our investigation predominantly revolves around GPTs, which have two available backend LLMs as of the manuscript submission date: GPT-3.5\footnote{GPTs can be indefinately tested with GPT-3.5-turbo model.} when designing and adjusting the GPTs, and GPT-4\footnote{Once released, GPTs can only be accessed via GPT-4 with ChatGPT premium.} when it is officially released. For a comparative analysis of jailbreak attacks, we replicate the queries on ChatGPT, also powered by GPT-4-turbo, to ascertain whether identical prompts yield similar jailbreak outcomes. Given the usage constraints of GPTs (limited to 40 queries every 3 hours at the time of this manuscript's submission), conducting a large-scale analysis is impractical. As a result, we limit our testing to 10 iterations for each of the 10 prompts designed for four prohibited scenarios (i.e., 100 tests per scenario) to minimize bias and ensure a more controlled study.

\noindent\textbf{Metrics.} Considering the variability in the content generated by the models, we undertake a manual inspection of each piece of content. This evaluation process involves marking a generation as a successful jailbreak attack based on specific criteria: (1) \textit{Relevance} - assessing whether the generated content is pertinent to the posed question; and (2) \textit{Content Quality} - determining if the content provides comprehensive and detailed instructions or explanations in response to the questions asked. This  ensures a thorough and accurate assessment of the effectiveness of the jailbreak attacks.

\subsection{Evaluation Results}

We follow the experiment settings to conduct the experiments, and the results are presented in  Table~\ref{tab:success_count_for_pattern_service}. Our findings indicate that \tool{} is remarkably effective in instigating jailbreak attacks across different scenarios. Notably, \tool{} demonstrates an average success rate of 64.3\% and 34.8\% on the four prohibited senarios over the GPT-3.5 and GPT-4 version of GPT instances, respectively. As a comparision, naive malicious question only achieve 3.0\% and 1.0\% of success rates over ChatGPT powered by the same models. This high success rate demonstrates  \tool{}'s capability to  achieve jailbreak by leveraging GPTs.

Analyzing the experimental results reveals several key insights. First, despite varying success rates across different models, the prohibited scenario of \textit{Privacy} is consistently the easiest to jailbreak, with an average success rate of 35.3\% in the four comparison groups. This finding aligns with conclusions from previous works that indicate varying difficulties in jailbreak scenarios. It suggests that while some content categories are more readily manipulated, others may require more sophisticated approaches for a successful jailbreak. Second, comparing naive injection over chatbots powered by GPT-3.5 and GPT-4 models, and GPT instances powered by these models, it is observed that GPT-4 powered chatbots and GPTs are more challenging to jailbreak. This can be attributed to the improved alignment in the training process of GPT-4. Finally, the repeatability of these results across multiple rounds underscores the consistency and reliability of \tool{} as a tool for probing the vulnerabilities of GPT models in the context of content policy violations.

\begin{table}[t]
\caption{Number and ratio of successful jailbreaking attempts for different models and scenarios. }
\label{tab:success_count_for_pattern_service}
\centering
\resizebox{\columnwidth}{!}{
%\fontsize{5.5}{10}\selectfont
\begin{tabular}{l||llll||l}
\toprule
\textbf{Pattern} & \textbf{Adult} & \textbf{Harmful} & \textbf{Privacy} & \textbf{Illegal} & \textbf{Average (\%)} \\
\midrule
Direct - GPT-3.5   & 1.0\%             & 2.0\%              & 6.0\%             & 3.0\%           & 3.0\%                     \\
Direct - GPT-4  & 0.0\%             & 0.0\%              & 1.0\%             & 3.0\%           & 1.0\%                      \\
GPTs - GPT-3.5   & 58.0\%             & 62.0\%              & 78.0\%             & 59.0\%           & 64.3\%    \\
GPTs - GPT-4   & 19.0\%             & 23.0\%              & 56.0\%             & 41.0\%           & 34.8\%    \\
\midrule
Average   & 19.3\%             & 21.8\%              & 35.3\%             & 26.5\%           & 25.7\%    \\
\bottomrule
\end{tabular}
}
\end{table}

\section{Conclusion and Future Works}\label{sec:future-works}

In this work, we unveil a novel approach for jailbreaking GPT models, termed RAG Poisoning. We developed \tool{} as a proof of concept to demonstrate the feasibility and effectiveness of this new attack method in real-world scenarios. Our preliminary results are quite revealing: {\tool{}} successfully executes jailbreak attacks across four distinct prohibited scenarios within GPTs, achieving a consistently high success rate. This achievement not only underscores the vulnerability of current GPT models to sophisticated attack strategies but also highlights the need for improvements in model resilience and security measures.

In the future, our research endeavors will branch out into several key directions, each aiming to further deepen our understanding and enhance the methodologies related to RAG Poisoning:

\noindent \textbf{Automated RAG Poisoning Development.} Currently, the knowledge base for GPT models is crafted manually, a process that is both time-intensive and potentially limited in scope. Our goal is to evolve this process into an automated pipeline. By doing so, we aim to streamline the generation of RAG content, thereby expanding the scale and diversity of the knowledge available for GPT models. This automation will not only improve efficiency but also enable the exploration of more complex and varied scenarios in RAG Poisoning.

\noindent \textbf{Enhancing RAG Poisoning Interpretability.} The current state of RAG Poisoning largely operates in a black-box nature, which poses challenges in understanding the underlying mechanisms and effects. Our objective is to transition this approach towards a more transparent, white-box model. This shift will allow for a deeper investigation into the causative factors behind successful jailbreak attacks orchestrated through RAG Poisoning. Unraveling these mechanisms, we gain critical insights into the vulnerabilities of LLMs and the dynamics of RAG interactions.

\noindent \textbf{Mitigation Strategies for RAG Poisoning.} Building upon the developments in automated RAG Poisoning and enhanced interpretability, our research will also focus on devising effective mitigation strategies against RAG Poisoning. This involves identifying and implementing safeguards to protect GPT models from being compromised by malicious RAG content. The integration of automated systems and a clearer understanding of RAG dynamics will be pivotal in developing robust defense mechanisms. These strategies will not only enhance the security and reliability of GPT models but also contribute to the broader field of AI safety and ethics.

\bibliographystyle{IEEEtran}
\bibliography{references}

\appendices

\end{document}